\documentclass[aps,prc,floatfix,groupedaddress,superscriptaddress,showpacs,preprint]{revtex4}
\usepackage{graphicx}
\usepackage{dcolumn}
\usepackage{bm}
\usepackage{amsmath}
\usepackage{amssymb}
\newcommand{\be}{\begin{equation}}
\newcommand{\ee}{\end{equation}}
\newcommand{\bc}{\begin{center}}
\newcommand{\ec}{\end{center}}
\newcommand{\bea}{\begin{eqnarray}}
\newcommand{\eea}{\end{eqnarray}}
\newcommand{\bean}{\begin{eqnarray*}}
\newcommand{\eean}{\end{eqnarray*}}
\newcommand{\ba}{\begin{array}{l}}
\newcommand{\ea}{\end{array}}
\newcommand{\Ds}{\displaystyle}

\newcommand*{\bpi}{{\bm{\pi}}}
\newcommand{\bA}{\mathbf{A}}
\newcommand{\bH}{\mathbf{H}}
\newcommand*{\bE}{\mathbf{E}}
\newcommand*{\vD}{\mathbf{D}}
\newcommand*{\bB}{\mathbf{B}}
\newcommand*{\vp}{\mathbf{p}}

\newcommand*{\bpsi}{\bm{\psi}}

\newcommand{\cH}{{\cal{H}}}
\newcommand{\cR}{{\cal{R}}}

\newcommand{\cE}{{\cal{E}}}

\newcommand{\bEvps}{\bE\cdot\bm{\psi}}

\newcommand{\divv}{{\rm div}}

\begin{document}

\title{The $s$-wave $\pi d$ scattering length from $\pi d$ atom using
effective field theory}
 \author{B.F. Irgaziev}
  \email{qcd@uzsci.net}
 \author{B.A. Fayzullaev}
  \email{bfayzullaev@nuuz.uzsci.net}

\affiliation{
    Theoretical Physics Department, National University of Uzbekistan  \\
 700174 Tashkent, Uzbekistan}

\date{\today}

\begin{abstract}
The $\pi^{-}d$ atom strong energy-level shift in the $1s$ state is derived
by using the effective field theory. Taking into account the large value of
radius of pionic deuterium and short radius of strong interaction between
pion and deuteron we have considered deuteron as particle described by vector
field. Pion is described by scalar field. To obtain non-relativistic
Hamiltonian for $\pi d$ system Foldy-Wouthuysen transformation has been
derived for the vector field. The strong interaction between pion and
deuteron has been taken at the zero-range approach. We have found the Deser
type formula for relation between the strong energy-level shift and the 
$s$-wave $\pi d$ scattering length.

\end{abstract}


\maketitle
\section{Introduction}

Recently, substantial progress has been made in developing Chiral
perturbation theory (ChPT) for pionic nucleon ($\pi N$) system
\cite{leut95,gass88,mojz98,fet98,fet00,bech99, Gasser:2002am}. In this context the precision
X-ray experiment on pionic hydrogen and pionic deuterium  carried out by
ETH Zurich-Neuch$\hat{\mbox{a}}$tel-PSI collaboration \cite{eth01}.  The $s-$wave scattering
$\pi N$ scattering lengths are important for testing of various theoretical
consideration like the Goldberger-Miyazawa-Oehme sum rule \cite{GMO}
determining the $\pi\pi N$ coupling constant. The sigma term, which is used
for lattice and ChPT calculation, is sensitively affected by the isoscalar
scattering length. The $\pi N$ and $\pi\,d$ scattering lengths are determined
as directly from the phase shift analysis so from the X-ray experiments on pionic atoms using the Deser formula \cite{deser54}.

The extraction of the $s$-wave scattering length from the X-ray experiments
gives errors less than the phase shift analysis. The accuracy of the modern
level of experimental analysis of the parameters of pionic hydrogen reached
by the PSI Collaboration is about 0.2\% for the strong energy shift and 1\%
for the width of the energy level of the ground state of pionic hydrogen
\cite{ana01}.  The NPE measurement \cite{haus98} of the pionic deuterium
atomic level strong shift yields for the $\pi d$ scattering length
$a_{\pi d}=[-0.0261 (\pm 0.0005) +i 0.0063 (\pm 0.0007)]m_{\pi}^{-1}$,
where $m_{\pi}$ is pion mass. Note an imaginary part of $\pi d$ scattering
length approximately in  four times less than its real part. Such results for 
$\pi^{-} d$ scattering can be explained by fact that the absorption channel
$\pi^{-}d\to nn$ and the radiative absorption channel $\pi^{-}d\to \gamma nn$
give very small contribution to the scattering length. Also
$\pi^{-}d\to \pi^{0}nn$ channel, which is opened at threshold, is suppressed
by the centrifugal barrier. Indeed, the 96\% of the deuteron state is
$^3S_{1}$ state, therefore according to Pauli principle $^3S_1$ is forbidden
for $nn$ system and  $^1S_0$ is not available if there is no spin flip.
In the series of papers \cite{petrov74,afnan74,mizut77,deloff01} the
determination of the pion-deuteron scattering length was considered by
solving the Faddeev equations. There has been remarkable recent
progress in developing the effective field theories to problems relevant to
the $\pi N$ \cite{rus00,ivan03} and $\pi d$  \cite{beane02} scattering length.
The main results which obtained from calculation of the $\pi d$
scattering length are following:
 (a) in all theoretical calculations isospin symmetry of the strong
interactions is assumed; (b) the $\pi^{-} d$ scattering length can be
extracted from the 1$s$ energy shift of pionic deuterium using the Deser
formula; (c) the electromagnetic contribution to the difference of the real
parts of the scattering length of $\pi^{+}$ and $\pi^{-}$ is of order of
the experimental error on the scattering length. The aim of this paper is
to establish the precise relation between the strong energy-level shift of
the $\pi^{-} d$ atom in the 1$s$ state, and strong $\pi d$ scattering length
using the effective field theory. In the present paper we show how to obtain
the nonrelativistic Hamiltonian for $\pi d$ system from the relativistic
equations of motion for the scalar and vector fields with taking into
account the electromagnetic interaction. The strong interaction is taken at
the zero-range approach.
\\ Here we use the unit system $\hbar=c=1$, therefore the fine-structure constant
is $\alpha=e^2$.

\section{Nonrelativistic limits for the scalar and vector fields}
\label{fields}
The problem of nonrelativistic limit description for fundamental particles
and their interactions may be solved in different ways. 
Although in all methods of nonrelativistic expansion the first terms of the Hamiltonians coincide, however the difference begins to arise at transition to the higher orders of expansion. The method of Foldy-Wouthuysen transformation
is one of the safest method for solving this problem \cite{bd65}.
This method was mainly used for spinor and scalar particles. In our task
we need in nonrelativistic Hamiltonian for massive vector field (deuteron is
considered as a fundamental particle) interacting with electromagnetic and
scalar (the pion) fields.
In this approach suggested by us the equations of motion are basic for producing 
nonrelativistic Hamiltonian. This Hamiltonian we obtain in the form of 
series over $1/M$, where $M$ is the mass of our vector particle.

We would like to point out some steps of derivation of the nonrelativistic limit Hamiltonians by means of the Foldy-Wouthuysen transformation \cite{bd65}. 
The method is based on the transformation of a relativistic equation of motion to the Schr\"odinger equation form.
For example, the relativistic equation of motion for free massive scalar field $\varphi$ is
\be
\left(\partial^2+m^2\right)\varphi=0,
\ee
or
\be\label{phi}
\partial_t^2\varphi=\left(\nabla^2-m^2\right)\varphi,
\ee
where $m$ is mass of the scalar particle, $\partial^2$ is the four-dimensional
D'Alambertian and $\partial_t={\partial}/{\partial t}$ (we use metric $g_{00}=+1, g_{11}=g_{22}=g_{33}=-1 $).
Let's introduce the "big" $\theta$ and the "small" $\chi $ components of
the field $\varphi$ as
\be
\theta=\frac{1}{2}(\varphi+\frac{i}{m}\partial_t\varphi),
\qquad\chi=\frac{1}{2}(\varphi-\frac{i}{m}\partial_t\varphi).
\ee
In the case $\varphi\sim e^{-imt} $ (i.e., when $\varphi$ corresponds to a particle) we have $\theta\sim \varphi,\,\chi\sim 0$,
and {\it vice versa} in the case $\varphi\sim e^{+imt}$ we have
$\theta\sim 0,\,\chi\sim \varphi$.
It is easy to see that
\be\ba
\Ds{i\partial_t\theta=m\theta-\frac{\nabla^2}{2m}(\theta+\chi)  }\\ \\
\Ds{i\partial_t\chi=-m\chi+\frac{\nabla^2}{2m}(\theta+\chi).}
\ea\ee
If we introduce a new two component function as
\be\Phi=\left(\begin{array}{c}\theta\\ \chi\end{array}\right),\ee
then we can rewrite Eq. (\ref{phi}) in the Schr\"odinder equation type
\be
i\partial_t\Phi=H\Phi,
\ee
where
\be
H=\eta\left(m+\frac{\vp^2}{2m}\right)+\rho\frac{\vp^2}{2m},
\ee
and
\be
\eta=\left(\begin{array}{cc} 1&0\\0&-1 \end{array}\right),\quad\rho=
\left(\begin{array}{cc} 0&1\\-1&0 \end{array}\right)
\ee
The matrix $\rho$ is that one which mixes the "big" and "small"  ("particle" and "antiparticle") components
of $\Phi$.
Using the unitary transformation
\be
\Phi^{\prime}=e^{iS}\Phi,\quad H^{\prime}=e^{iS}He^{-iS},
\ee
searching the operator $S$ in the form
\be S=\eta\rho\lambda\,,\ee
and demanding removing of terms of odd order in $\rho$ in the Hamiltonian
we
find the operator $\lambda$ of the Foldy-Wouthuysen transformation to be
\be
i\tan(2\lambda)=\frac{\vp^2}{m^2+p_0^2},
\ee
and the new Hamiltonian
\be
H^{\prime}=\eta p_0=\eta\sqrt{m^2+\vp^2}.
\ee
So, our new Schr\"odinger equation $i\partial_t\Phi^\prime=H^\prime\Phi^\prime$ is free of mixing terms.


Now we include the electromagnetic field. The equation of motion is
\be
\left( D^2+m^2 \right)\varphi=0,\quad {\rm or }\quad D_0^2\varphi=
\left(\vD^2-m^2\right) \varphi,
\ee
where we have used the standard notations
\be D_\mu=\partial_\mu-ieA_\mu,\quad D_0=\partial_0-ieA_0,\quad \vD=
-\nabla-ie\bA,\quad \bpi=i\vD=\vp+e\bA,\ee
and $A_\mu$ is the electromagnetic potential, $\bpi$ is generalized momentum.
Determining the "big" and "small" components of $\varphi$ by equations
\be
\theta=\varphi+\frac{i}{m}D_0\varphi,\quad \chi=\varphi-\frac{i}{m}D_0\varphi,
\ee
we receive
\be
iD_0\theta=m\theta-\frac{\vD^2}{2m}\left( \theta+\chi\right), \quad
iD_0\chi=-m\chi+\frac{\vD^2}{2m}\left( \theta+\chi\right),
\ee
or
\be
i\partial_t\Phi=H\Phi,
\label{ef}\ee
where
\be
H=m\eta -eA_0- \eta\frac{\vD^2}{2m}-\rho\frac{\vD^2}{2m}=m\eta -eA_0+ \eta
\frac{\bpi^2}{2m}+\rho\frac{\bpi^2}{2m}.
\ee
This Hamiltonian we should transform to separate the states with the positive
energy from that of negative ones. Substituting
\be
\Phi=e^{-iS}\Phi^{\prime}
\ee
into Eq. (\ref{ef}) we obtain
\be
i\partial_t\Phi^{\prime}=H^{\prime}\Phi^{\prime},
\label{efp}\ee
where
\be
H^{\prime}=e^{iS}\left( H-\partial_t S\right) e^{-iS}.
\ee
Using standard expansion formula
\be
e^{iA}Be^{-iA}=B+i[A,B]+\frac{i^2}{2}[A,[A,B]]+\frac{i^3}{3}[A,[A,[A,B]]]+...
\ee
and taking into account that $S\sim 1/m$, which follows from the free case, we get
\bea
H^{\prime}&=&H+i[S,H]-\frac{1}{2} [S,[S,H]]-\frac{i}{6}[S,[S,[S,H]]]
\nonumber\\
&+&\frac{1}{24}[S,[S,[S,[S,H]]]]
-\partial_t S-\frac{1}{2}[S,\partial_t S]+\frac{1}{6}
[S,[S,\partial_t S]],
\eea
up to the accuracy $O(1/m^4)$ \cite{bd65}.

Let us use the following notation
\be
H=m\eta -eA_0- \eta\frac{\vD^2}{2m}-\rho\frac{\vD^2}{2m}=\eta m+\cE+\cR,
\ee
where
\be\cE=-eA_0-\eta\vD^2/2m=-eA_0+\eta\frac{\bpi^2}{2m},\qquad\cR=-
\rho\vD^2/2m=\rho\frac{\bpi^2}{2m}.\ee
We suppose that the Coulomb energy has the same order of magnitude as the kinetic energy.
It is usual situation  for nonrelativistic case.
Our aim is to eliminate terms with odd degrees of $\cR$. This work may be done only step-by-step.
After some steps we obtain the expression which is free of odd degrees of $\cR$ with sufficient accuracy (see \cite{bd65}):
\be
H^{\prime}= \eta\left( m+\frac{1}{2m}\cR^2-\frac{1}{8m^3}\cR^4 \right)+
\cE-\frac{1}{8m^2}[\cR,[\cR,\cE]]-\frac{i}{8m^2}[\cR,\partial_t\cR]+O(1/m^4).
\ee

It is not difficult to show that
\be
[\cR,\cE]+i\partial_t \cR=-\frac{ie}{m}\rho\bpi\cdot\bE.
\ee
So, we have the transformed Hamiltonian up to order $O(1/m^4)$ in the form
\be
H^{\prime}= \eta\left( m+\frac{\bpi^2}{2m}-\frac{\bpi^4}{8m^3}\right)-
eA_0-\frac{ie}{16m^4}[\bpi^2,\bpi\cdot\bE].
\ee
Here we have no terms mixing the "small" and the "big" components.
This Hamiltonian includes terms of order $(v/c)^2$.


Now we can proceed to a massive vector field theory.
A massive  particle of spin 1  has three degrees of freedom, so , it may be considered
as spatial part of a four vector $\psi_\mu$.
In Ref. \cite{schw40} a general theory describing particles of unit spin
and arbitrary magnetic moment was developed and applied to the motion of
such particles in the electromagnetic field. For our purpose we use the
equations of motion given in Ref. \cite{wen,pauli}. The massive
four-dimensional vector field $\psi_\nu$ is described by equations
\be
{\psi}_{\mu\nu}=D_\mu{\psi}_\nu-D_\nu{\psi}_\mu;\qquad
D^\mu{\psi}_{\mu\nu}+M^2{\psi}_\nu-ie\kappa F_{\nu\mu}{\psi}^\mu=0,
\label{epsi}\ee
where $M$ is mass of the vector particle,
\be F_{\nu\mu}=\partial_\mu A_\nu-\partial_\nu A_\mu \ee
is the strength
tensor of the electromagnetic field,
\be D_\mu=\partial_\mu-ieA_\mu \ee
is the electromagnetic covariant derivative and $\kappa$ is the anomaly part of
magnetic moment of the particle. Acting on the second equation by $D^\nu$ we receive
the following constraint:
\be\label{cons}
D^\mu{\psi}_\mu=-\frac{ie}{2M^2}F^{\mu\nu}{\psi}_{\mu\nu}+
\frac{ie\kappa}{M^2}D^\mu(F_{\mu\nu}{\psi}^\nu).
\ee
This constraint reflects the fact that in the Lorentz covariant description
of a massive vector field with three degrees of freedom we have
used a four vector $\psi_\mu$, so one of its components must be excluded.

Let us rewrite the Eq.(\ref{epsi}) as
\be
D^2{\psi}_\nu-D^\mu D_\nu{\psi}_\mu+M^2{\psi}_\nu-ie\kappa
F_{\nu\mu}{\psi}^\mu=0,
\ee
or,
\be
D^2{\psi}_\nu-D_\nu D^\mu{\psi}_\mu+M^2{\psi}_\nu-ie(1+\kappa)
F_{\nu\mu}{\psi}^\mu=0,
\ee
Using the constraint (\ref{cons}) we get
\be
D_0^2{\psi}_\nu=\left(\vD^2-M^2 \right){\psi}_\nu +ie(1+\kappa)
F_{\nu\mu}{\psi}^\mu-\frac{ie}{2M^2}\left[D_\nu (F^{\mu\nu}
{\psi}_{\mu\nu})-2\kappa D_\nu D^\mu(F_{\mu\lambda}\psi^\lambda) \right].
\label{e2}\ee
When it needed we should replace the zero component of $\psi_\nu$ through  its spatial part (cf. Eq. (\ref{epsi})  ):
\be
\psi_0=\frac{1}{M^2-\vD^2}\left[ ie(1+\kappa)\bEvps+D_0D^i\psi_i \right],
\label{psi0}\ee
where $\psi_i$ are spatial   components of $\bpsi$. It is not difficult to convince that
Eqs. (\ref{cons}) and (\ref{psi0}) coincide. Further we will write out equations for the spatial part
of $\psi_\mu$ only.

Let us to introduce the following notations
\be
\zeta_i={\psi}_i+\frac{i}{M}D_0{\psi}_i\,,\qquad \chi_i={\psi}_i-
\frac{i}{M}D_0{\psi}_i.
\ee
Then
\bea
iD_0\zeta_i&=&iD_0{\psi}_i-\frac{1}{M}D_0^2{\psi}_i=M\zeta_i-
\frac{\vD^2}{M}{\psi}_i-\frac{ie(1+\kappa)}{M}F_{i\lambda}\psi^\lambda +
O(\frac{1}{M^3}),\\
iD_0\chi_i&=&iD_0{\psi}_i+\frac{1}{M}D_0^2{\psi}_i=-M\chi_i+
\frac{\vD^2}{M}{\psi}_i+\frac{ie(1+\kappa)}{M}F_{i\lambda}
\psi^\lambda-O(\frac{1}{M^3}).
\eea
Supposing
\be
\Psi_i=\left(\begin{array}{c}\zeta_i\\ \chi_i\end{array}\right)
\ee
we can rewrite the last system of equations as
\be
i\partial_0\Psi_i=H_i^j\Psi_j.
\ee
Nonrelativistic equations for components of $\Psi_i$ may be written as
\bea
i\partial_0\zeta_i&=&(M-eA_0)\zeta_i-\frac{\vD^2}{2M}(\zeta_i+\chi_i)+
\frac{ie(1+\kappa)}{2M}\left(\bB\times(\bm{\zeta}+\bm{\chi})\right)_i+
O(\frac{1}{M^3})\, , \\
i\partial_0\chi_i&=&-(M+eA_0)\chi_i+\frac{\vD^2}{2M}
(\zeta_i+\chi_i)-\frac{ie(1+\kappa)}{2M}\left(\bB\times
(\bm{\zeta}+\bm{\chi})\right)_i+O(\frac{1}{M^3})\, .
\eea
In the two-component notations we have
\be
i\partial_0\Psi_i=\left( -eA_0+M\eta-\frac{\vD^2}{2M}(\eta+\rho) \right)
\Psi_i+\frac{ie(1+\kappa)}{2M}(\eta+\rho)(\bB\times\bm{\Psi})_i+
O(\frac{1}{M^3}).
\ee
The situation is like to the scalar case except the term with
magnetic field.
Denoting
\be
H_{ij}=M\eta\delta_{ij}+\cE_{ij}+\cR_{ij}
\label{eh}\ee
we obtain
\bea
\cE_{ij}&=&\left(-eA_0-\eta\frac{\vD^2}{2M} \right) \delta_{ij}+
\frac{ie(1+\kappa)}{2M}\eta\varepsilon_{ikj}B_k,\label{Eij}\\
\cR_{ij}&=&-\rho\frac{\vD^2}{2M}\delta_{ij}+\frac{ie(1+\kappa)}
{2M}\rho\varepsilon_{ikj}B_k.\label{Rij}
\eea
Let's at the first step restrict by terms of $1/M$ order.
In this case we have
\be
i\partial_0\Psi_i=\eta\left( M+\frac{\bpi^2}{2M}\right)\Psi_i+
\frac{ie(1+\kappa)}{2M}\eta(\bB\times\bm{\Psi})_i-eA_0\Psi_i.
\ee
For the Hamiltonian density we receive
\be
\cH=\Psi^\dagger_{di}\Big[\eta\left( M+\frac{\bpi^2}{2M}\right)-
eA_0\Big]\Psi_{di} + \frac{ie(1+\kappa)}{2M}\bm{\Psi}^\dagger_d\eta\cdot\bB
\times\bm{\Psi}_d,
\ee
where $\bm{\Psi}_d$ denotes the non-relativistic vector field operator,
$\Psi_{di}$ are components of $\bm{\Psi}_d$
This is a nonrelativistic Hamiltonian where the spin (magnetic moment) of
$\psi$-field is taken into account.

The procedure of getting the next terms is the same as for scalar field.
The result of the Hamiltonian without the terms mixing particle-antiparticle
is:
\be
H^{\prime}= \eta\left( M+\frac{1}{2M}\cR^2-\frac{1}{8M^3}\cR^4 \right)+
\cE-\frac{1}{8M^2}[\cR,[\cR,\cE]]-\frac{i}{8M^2}[\cR,\partial_t\cR] .
\label{hpp}\ee
Here it is assumed, that all terms enter with the additional tensor structures which
are defined  in Eqs.  (\ref{eh}) - (\ref{Rij})

The terms up to order $1/M^4$ are included into Eq. (\ref{hpp}). In this work we
will restrict ourself by terms of order  $1/M^3$ only omitting all terms of higher accuracy.
Therefore for the vector field we obtain the Hamiltonian density
\be
\cH^{\prime}=\Psi^\dagger_{di}\Big[\eta\left( M+\frac{\bpi^2}{2M}-
\frac{\vp^4}{8M^3}\right)-eA_0\Big]\Psi_{di}
+\frac{ie(1+\kappa)}{2M}
\bm{\Psi}^\dagger_d\eta\cdot\bB\times\bm{\Psi}_d.
\label{ech}\ee

Now we can write the total Hamiltonian including into consideration other
fields that means pion and electromagnetic fields. Also we include into
the density of the Hamiltonian the strong $\pi d$ interaction which is
taken in the zero range approach
\be\label{str}
\cH_{S}=-d_{\pi d}\bm{\Psi}^\dagger_d\Phi_\pi^\dagger\bm{\Psi}_d\Phi_\pi,
\ee
where $\Phi_\pi$ denotes the non-relativistic field operator of pion and
$d_{\pi d}$ is the coupling constant which does not depend on the spin
projection of deuteron.
The electromagnetic Lagrangian is
\be
L_{em}=-\frac{1}{4}F_{\mu\nu}^2=\frac{1}{2}(\bE^2-\bB^2)=
-\bE\cdot\partial_0\bA+A_0{\rm div}\bE-\frac{1}{2} (\bE^2+\bB^2),
\ee
i.e.,
\be \cH_{em}=\frac{1}{2} (\bE^2+\bB^2)\ee
is the (density) of the electromagnetic Hamiltonian  and $A_0$ is a
Lagrange multiplier (it is not a dynamical variable).

Taking into account the Lagrange multiplier term
we get the total Hamiltonian density in form
\bea
\cH&=&\Psi^\dagger_{di}\left[\eta\left( M+\frac{\bpi^2}{2M}-
\frac{\vp^4}{8M^3}\right)+eA_0
\right]\Psi_{di} - \frac{ie(1+\kappa)}{2M}\bm{\Psi}^\dagger_d\eta\cdot\bB
\times\bm{\Psi}_d\nonumber\\
&+&\frac{1}{2}(\bE^2+\bB^2)-}
A_0{\rm div}\bE+\Ds{\Phi_\pi^\dagger \left[\eta\left( m+\frac{\bpi^2}{2m}-
\frac{\vp^4}{8m^3}\right)-
eA_0\right]\Phi_\pi
\label{total} \\
&-&d_{\pi d}\bm{\Psi}^\dagger_d\Phi_\pi^\dagger\bm{\Psi}_d
\Phi_\pi. \nonumber
\eea
In the gauge $\divv \bA=0$ we have $\divv \bE=-\Delta A_0$ (due to $\bE=-\nabla A_0-\partial_t \bA$), so, the Coulombic
part of our Hamiltonian is
$$ A_0\Delta A_0+\frac{1}{2} (\nabla A_0)^2+\rho A_0,\qquad \rho=
e\left(\bm{\Psi}^\dagger_d\cdot\bm{\Psi}_d-\Phi_\pi^\dagger
\Phi_\pi \right) $$
Excluding $A_0$  and neglecting the spin-flip part of the density Hamiltonian
(\ref{total}) we obtain the canonical Hamiltonian for the
non-relativistic $\pi d$ system as
\bea\label{Hamiltonian}
\bH=\bH_{\rm 0}+\bH_{\rm C}+\bH_{\rm R}
+\bH_\gamma+\bH_{\rm S}
=\bH_{\rm 0}+\bH_{\rm C}+{\bf V}\, ,
\eea
where $\bH_0$ is the free Hamiltonian describing
non-relativistic pions and deuterons. Further,
$\bH_\Gamma=\int d^3{\bf x}\, \cH_\Gamma,
~\Gamma={\rm C,R,}\gamma,{\rm S}$, and
\bea\label{coul}
\cH_{\rm C}&=&
e^2\,(\bm{\Psi}^\dagger_d\bm{\Psi}_d)\,\triangle^{-1}\,
(\Phi^\dagger_\pi\Phi_\pi)\, ,
\\[2mm]
\cH_{\rm R}&=&
-\bm{\Psi}^\dagger_d\, \frac{\nabla^4}{8M^3} \,\bm{\Psi}_d
-\Phi^\dagger_\pi, \frac{\nabla^4}{8m^3} \,\Phi_\pi\, ,
\label{rel}\\[2mm]
\cH_\gamma&=&
-\frac{ie}{2M}\,\bm{\Psi}^\dagger_d\, (\nabla{\bA}+{\bA}\nabla)\,
\bm{\Psi}_d +\frac{ie}{2m}\,
\Phi^\dagger_\pi\, (\nabla{\bA}+{\bA}\nabla)\, \Phi_\pi\, ,
\label{mag}\\[2mm]
\cH_{\rm S}&=&
-d_{\pi d}\,\bm{\Psi}^\dagger_d\bm{\Psi}_d\,\Phi^\dagger_\pi\Phi_\pi\, .
\label{str1}
\eea

\section{Energy shift of pionic deuterium atom}

We consider the problem of $s$-state energy shift according to  the perturbation theory.
Such analysis was performed for the pionic hydrogen in Ref. \cite{rus00}.  Let
${\bf H}_0+{\bf H}_{\rm C}$ is the unperturbed Hamiltonian,
whereas ${\bf V}$ is considered as a perturbation.
The ground-state solution of the unperturbed Schr\"odinger equation
in the center of mass (CM) system frame  \\
$(\tilde E_1-{\bf H}_0-{\bf H}_{\rm C})|\Psi_0({\bf 0})\rangle=0$, with
$\tilde E_1=M+m+E_1$, is given by
\be\label{psi_0}
|\Psi_0({\bf 0})\rangle=\int\frac{d^3{\bf p}}{(2\pi)^3}\,\,\Psi_0({\bf p})
\,\, b^\dagger({\bf p})\,a^\dagger(-{\bf p})\, |0\rangle\, ,
\ee
where $a^\dagger({\bf p})$ and $b^\dagger({\bf p})$
denote creation operators for non-relativistic $\pi^-$ and deuteron
acting on the Fock space vacuum, and $\Psi_0({\bf p})$ stands for
the non-relativistic Coulomb wave function of $1S$ state in the momentum space,
$E_1$ is the non-relativistic binding energy of pionic deuterium.
We normalize the ground-state wave function as
\be
\int\,d\,^3\,\rm{p}|\Psi_0({\bf p})|^2=1.
\ee

We are going to evaluate the energy-level shift of the ground
state due to the perturbation Hamiltonian ${\bf V}$ as in Ref. \cite{rus00}.
Let us define the free and Coulomb Green operators by the expressions
${\bf G}_0(z)=(z-{\bf H}_0)^{-1}$ and
${\bf G}(z)=(z-{\bf H}_0-{\bf H}_{\rm C})^{-1}$, respectively.
Further, we define the ``Coulomb-pole removed'' Green function as
$\hat {\bf G}(z)={\bf G}(z)(1-{\bf \Pi})$, where ${\bf \Pi}$ denotes the
projector onto the Coulomb ground state $\Psi_0$ (\ref{psi_0}).
The $\pi d$ scattering states in the sector with the total charge $0$ are
defined as
$|{\bf P},{\bf p}\rangle=b^\dagger({\bf p}_1)\,
a^\dagger({\bf p}_2)\, |0\rangle$ (${\bf p}_1$ and ${\bf p}_2$ denote momenta of deuteron and pion, respectively).
The CM and relative momenta are defined by
${\bf P}={\bf p}_1+{\bf p}_2$,
${\bf p}=(m\,{\bf p}_1-M\,{\bf p}_2)/(M+m)$.
We remove the CM momenta from the matrix elements of any operator
${\bf R}(z)$ by introducing the notation
\be\label{CM}
\langle {\bf P},{\bf q}|{\bf R}(z)|
{\bf 0},{\bf p}\rangle=
(2\pi)^3\delta^3({\bf P})\,
({\bf q}|{\bf r}(z)|{\bf p})\, .
\ee

The ``Coulomb-pole removed'' transition operator satisfies the equation
\be\label{pole_C}
{\bf M}(z)={\bf V}+{\bf V}\hat {\bf G}(z){\bf M}(z)\, .
\ee

According to the Feshbach's formalism \cite{fesh58} the scattering operator ${\bf T}(z)$
develops the pole at
$z=\bar z$ where $\bar z$ is the solution of the following equation
\be\label{pole_z}
\bar z-\tilde E_1-(\Psi_0|{\bf m}(\bar z)|\Psi_0)=0\, ,
\ee
where $({\bf p}|\Psi_0)=\Psi_0({\bf p})$ and ${\bf m}(z)$ is
related to ${\bf M}(z)$ through the definitions~(\ref{CM}).

In order to get the shift of the ground-state energy,
the quantity ${\bf m}(z)$ is calculated perturbatively
from Eq.~(\ref{pole_C}) by the iteration series at accuracy
$O(\alpha^4)$. Using the explicit expression of ${\bf V}$ given
by Eq. (\ref{Hamiltonian}),
replacing $\hat {\bf G}(z)$ by ${\bf G}_0(z)$ whenever possible,
and retaining only those terms that contribute at the accuracy we are
working, the operator ${\bf M}(z)$ can be written in the form
${\bf M}(z)={\bf U}(z)+{\bf W}(z)$, where
\bea\label{uw}
{\bf U}(z)&=&{\bf H}_{\rm R}
+{\bf H}_\gamma {\bf G}_0(z){\bf H}_\gamma
\, ,
\nonumber\\[2mm]
{\bf W}(z)&=&{\bf H}_{\rm S}+{\bf H}_{\rm S}\hat {\bf G}(z){\bf H}_{\rm S}
\, ,
\eea

At the accuracy $O(\alpha^4)$, the energy of the bound state is equal to
$\bar z=\tilde E_1+\Delta E^{\rm em}_1+\epsilon_{1s}$, where
\be\label{em_shift}
\Delta E^{\rm em}_1
={\rm Re}\, (\Psi_0|{\bf u}(\tilde E_1)|\Psi_0)+E^{\rm vac}\, ,
\quad
\epsilon_{1s}
={\rm Re}\, (\Psi_0|{\bf w}(\tilde E_1)|\Psi_0)\, .
\ee
Here ${\bf u}(z)$, ${\bf w}(z)$ are related to
${\bf U}(z)$, ${\bf W}(z)$ through the definitions (\ref{CM}) and
$E^{\rm vac}$ stands for the contribution due to the electron vacuum
polarization which is added ``by hand''.
The results of calculations for the matrix elements determining the
energy-label shift are the same as in Ref.~\cite{rus00}:
\bea\label{Re}
&&\hspace*{-0.6cm}
{\rm Re}\, (\Psi_0|{\bf u}(\tilde E_1)|\Psi_0)=
-\frac{5}{8}\,\alpha^4\mu_{\pi d}^4\,
\frac{M^3+m^3}{M^3m^3}
-\frac{\alpha^4\mu_{\pi d}^3}{M m}\, ,
\\[2mm]
&&\hspace*{-0.6cm}
{\rm Re}\, (\Psi_0|{\bf w}(\tilde E_1)|\Psi_0)=
\frac{\alpha^3\mu_{\pi d}^3}{\pi}\, \biggl[
-d_{\pi d}+d_{\pi d}^2\, (\,\xi+\frac{\alpha\mu_{\pi d}^2}{\pi}\, (\ln\alpha-1))
\biggr]\, ,
\nonumber
\eea
where $\mu_{\pi d}$ is the reduced mass of the $\pi d$ system, and
\bea\label{xi}
\xi&=&\frac{\alpha\mu_{\pi d}^2}{2\pi}\biggl\{
(\mu^2)^{d-3}\biggl(\frac{1}{d-3}-\Gamma'(1)-\ln 4\pi\biggr)
+\ln\frac{(2\mu_{\pi d})^2}{\mu^2}-1\biggr\}\, .
\eea
Here $d$ and $\mu$ denote the dimension of space and the scale of the
dimensional regularization used as in Ref.~\cite{rus00}, respectively.

The energy shift ( order $\alpha^2$ ) due to the vacuum polarization contribution is given by the well-known
expression \cite{rus00}
\bea\label{vac-pol}
&&\hspace*{-0.3cm}
E^{\rm vac}
=-\frac{\alpha^3\mu_{\pi d}}{3\pi}\,\eta^2\Phi(\eta),\quad
\eta=\frac{\alpha\mu_{\pi d}}{m_e}\, ,
\\[2mm]
&&\hspace*{-0.3cm}
\eta^2\Phi(\eta)=\frac{1}{\eta^3}\biggl(2\pi-4\eta+\frac{3\pi}{2}\,\eta^2
-\frac{11}{3}\,\eta^3\biggr)
+\frac{2\eta^4-\eta^2-4}{\eta^3\sqrt{\eta^2-1}}\,
\ln(\eta+\sqrt{\eta^2-1})\, ,
\nonumber
\eea
where $m_e$ denotes the electron mass.

The calculation of the electromagnetic energy-level shift is now complete.

We present our results in the convenient form
\be\label{conv}
\tilde E_1+\Delta E_1^{\rm em}=E^{\rm KG}+E^{\rm vac}+
E^{\rm rel}\, ,
\ee
where
\bea\label{em1}
E^{\rm KG}&=&
-\frac{1}{2}\, \mu_{\pi d} \alpha^2\,\biggl(1+\frac{5\alpha^2}{4}\biggr)\, ,
\\[2mm]
E^{\rm rel}&=&\frac{7}{8}\,\alpha^4\frac{\mu_{\pi d}^3}{M m}\, ,
\eea
and $E^{\rm vac}$ is given by Eq.~(\ref{vac-pol}). In Table~1
we present various contributions to the electromagnetic energy-level shift of
$\pi^- p$ and $\pi^- d$ atoms.  As can be seen
from Table~1, the calculated values of the electromagnetic
shifts for $\pi^- p$ and $\pi^- d$ atoms are close to the each other.

\begin{figure}[t]
{\small
{\bf Table 1.}
Contributions to the electromagnetic binding energy of the $\pi^- p$
and $\pi^- d$ atoms (eV).

\vspace*{.3cm}

\noindent
\begin{tabular}{ l l r r }
\hline\hline
Type of contribution & Notation & \hspace*{1.2cm} $\pi^- p$ & \hspace*{1.2cm} $\pi^- d$      \\
\hline
Point Coulomb, KG equation              &$E^{\rm KG}$   & $-3235.156$ & $-3459.0$    \\
Vacuum polarization, order $\alpha^2$   &$E^{\rm vac}$  & $-3.241$    & $-3.732$   \\
Relativistic recoil                     &$E^{\rm rel}$  & $0.034$     & $0.021$      \\
\hline\hline
\end{tabular}
}

\vspace*{.5cm}

\end{figure}

\section{$S$-wave scattering length and the Deser's type formula}

In order to complete the calculation of the strong energy-level shift, one
has to match at the accuracy $O(\alpha)$ the particular combination of
the non-relativistic coupling $d_{\pi d}$. Therefore  we consider
the scattering operator
\bea\label{TR}
&&{\bf T}_{\rm R}(z)={\bf V}_{\rm R}+{\bf V}_{\rm R}{\bf G}_{\rm R}(z)
{\bf T}_{\rm R}(z)\, ,
\nonumber\\[2mm]
&&{\bf V}_{\rm R}={\bf H}_{\rm C}+{\bf H}_\gamma+{\bf H}_{\rm S}\, ,\quad
{\bf G}_{\rm R}(z)=(z-{\bf H}_0-{\bf H}_{\rm R})^{-1}\, .
\eea

In the scattering operator ${\bf T}_{\rm R}(z)$, all kinematical insertions
contained in ${\bf H}_{\rm R}$ are summed up in the external lines
(see~\cite{ant00} for the details). We calculate the matrix element of the
scattering operator ${\bf T}_{\rm R}(z)$ between the $\pi^-d$ states at
$O(\alpha)$. After removing the CM momentum, the spin-nonflip part of
this matrix element on energy shell is equal to
\bea\label{T-alpha}
&&\hspace*{-0.9cm}
({\bf q}|{\bf t}_{{\rm R}}(z)|{\bf p})=
-\frac{4\pi\alpha}{|{\bf q}-{\bf p}|^2}-\frac{4\pi\alpha}{4Mm}\,
\frac{({\bf q}+{\bf p})^2}{|{\bf q}-{\bf p}|^2}
+{\rm e}^{2i\alpha\theta_{\rm C}(|{\bf p}|)}\,
({\bf q}|\bar {\bf t}_{{\rm R}}(z)|{\bf p})\, ,
\nonumber\\
&&
\eea
where the (divergent) Coulomb phase in the
dimensional regularization scheme is given by
\be\label{phase}
\theta_{\rm C}(|{\bf p}|)=\frac{\mu_{\pi d}}{|{\bf p}|}\,
\mu^{d-3}\,\biggl(\frac{1}{d-3}-\frac{1}{2}(\Gamma'(1)+\ln 4\pi)\biggr)
+\frac{\mu_{\pi d}}{|{\bf p}|}\,\ln\frac{2|{\bf p}|}{\mu}\, ,
\ee
and
\be\label{T-threshold}
{\rm Re}\,({\bf q}|\bar {\bf t}_{\rm R}(z)|{\bf p})=
-\frac{\pi\alpha\mu_{\pi d} d_{\pi d}}{|{\bf p}|}
+\frac{\alpha\mu_{\pi d}^2d_{\pi d}^2}{\pi}\, \ln\frac{|{\bf p}|}{\mu_{\pi d}}
-d_{\pi d}
+d_{\pi d}^2\xi+\cdots\, ,
\ee
where ellipses stand for the terms that vanish at threshold. The first two terms
in Eq. (\ref{T-threshold}) are arisen due to the electromagnetic interactions
between deuteron and pion, the sum of two next terms is amplitude of
scattering due to the strong interaction at threshold.

Therefore for the regular part of the $s$-wave scattering length we have
\be\label{scat-A}
-\frac{2\pi}{\mu_{\pi d}}{\cal A}_{\pi d}=
-d_{\pi d}+d_{\pi d}^2\xi.
\ee
Finally compare Eq.~(\ref{scat-A}) with Eq.~(\ref{Re}) where we take into account all
terms at accuracy $O(\alpha^3)$ we obtain the relation between the regular
part of the $s$-wave $\pi d$ scattering length and the strong energy-level
shift of $\pi^- d$ atom:
\be\label{strong}
\epsilon_{1s}=-2\alpha^3\mu_{\pi d}^2\, {\cal A}_{\pi d}\, ,
\ee
where the ultraviolet divergence contained in the quantity $\xi$, has been canceled.

\section{Discussion and conclusion}
From the measurement of the $3p - 1s$ X-ray transition of pionic deuterium
the $\pi^- d$ scattering length has been determined.
The strong energy-level shift can be defined by
\be\label{x-ray}
\epsilon_{1s}=E_{3p-1s}^{mag}-E_{3p-1s}^{meas}\,,
\ee
where $E_{3p-1s}^{mag}$ is the electromagnetic transition energy calculated
in the absence of the strong interaction \cite{sigg96} (the strong interaction
in the $3p$ state is negligible), $E_{3p-1s}^{meas}$ is the measured transition energy.
It is clear that on the extracted significance
of $\epsilon_{1s}$ energy shift the contributions to the electromagnetic energy
$E_{3p-1s}^{mag}$ are affected. According to our suggested model the contribution
of deuteron magnetic moment would be differed from the results of the standard
model using for calculation of the electromagnetic transition. Therefore it will
be useful to estimate this contribution.

In order to combine the results from pionic hydrogen and pionic deuterium
the establish of relation between ${\cal A}_{\pi d}$ and $\pi N$ isoscalar and
isovector lengths $b_0$ and $b_1$ is the important problem. This relation
can be expressed as
\be\label{A-high}
{\cal A}_{\pi d}=\frac{1+m/m_N}{1+m/M}2b_0+{\cal A}_{\pi d}^{\mbox{\small(higher order)}}\, ,
\ee
where the first terms is the scattering amplitude in the impulse approximation,
$2b_0$ is the sum of the amplitude for $\pi^- p$ and $\pi^- n$ elastic scattering,
$m_N$ is nucleon mass. The second term is dominant and it includes all remaining
(higher order) contributions. The results presented in Ref.~\cite{eth01} shows
that the $s$-wave double scattering gives the most contribution while the multiple
scattering and form factor correction are much smaller. It is well-known that the
second-order formula for $\pi d$ scattering length is described by (cf. Ref.~\cite{ericson})
\be\label{A-second}
{\cal A}_{\pi d}=\frac{2\mu_{\pi d}}{m}\left[\tilde{b}_0+(\tilde{b}_0^2-
\tilde{b}_1^2)\left\langle\frac{1}{r}\right\rangle\right]\, ,
\ee
where the expectation value of $1/r$ is taken with respect to the deuteron wave
function and $\tilde{b}_j=(1+m/m_N)b_0$. The calculations of the expectation value
performed by the various $NN$ potentials gave 20\% of discrepancy for its value
\cite{deloff01}. That means it is necessary to calculate the expectation value
with potential which has the right behavior at the small distance. 
We note that absolute value of ${\cal A}_{\pi d}$ (Eq. (\ref{A-second})), calculated with use of various $NN$ potentials leads to the larger value than at the solution of the Faddeev equation with the same potentials. 
On the basis of the solution the Faddeev equation received  in
Ref.~\cite{deloff01} the formula (\ref{A-second}) can be extended
\be\label{A-extend}
{\cal A}_{\pi d}=\frac{2\mu_{\pi d}}{m}\left[\tilde{b}_0+(\tilde{b}_0^2-
\tilde{b}_1^2)\left\langle\frac{e^{-\kappa r}}{r}\right\rangle +
(\tilde{b}_0+\tilde{b}_1)^2(\tilde{b}_0-2\tilde{b}_1)\left\langle
\frac{e^{-2\kappa r}}{r^2}\right\rangle\right] \, ,
\ee
where $\kappa=\sqrt{2\mu_{np}\epsilon_d}$ and $\epsilon_d$ is the binding energy
of the deuteron, $\mu_{np}$ is reduced mass for proton and neutron. In contrast to
Eq.~(\ref{A-second}) the binding energy correction of deuteron and the other
higher order corrections are taken into account in Eq.~(\ref{A-extend}).

At the end we would like to emphasize while we have not taken into account
the important parameters of deuteron and pion as their sizes, but  it is possible
to construct the non-relativistic Lagrangian, like in \cite{rus00}, which includes
these parameters through coupling constants.

We are grateful to Prof. J.~Gasser  and Dr. A.~Rusetsky for the current interest in the
work and useful suggestions.
This work was supported by the Swiss National Science
Foundation, SCOPES 2000–2003, Project No. 7UZPJ65677.

\end{document}